\documentclass[11pt]{article}

\usepackage{amsmath}
\usepackage{amsthm}
\usepackage{titlesec}
\usepackage{indentfirst}
\theoremstyle{definition}
\theoremstyle{plain}\newtheorem{Th}{Theorem}
\theoremstyle{definition}
\theoremstyle{definition}
\theoremstyle{plain}
\theoremstyle{plain}\newtheorem{Co}[Th]{Corollary}
\theoremstyle{plain} \textwidth 155mm
\begin{document}

\title{{\bf Probabilistic cloning with
supplementary information contained in the quantum states of two
auxiliary systems\thanks{This research is supported by the
National Natural Science Foundation (Nos. 90303024, 60573006), the
Research Foundation for the Doctoral Program of Higher School of
Ministry of Education (No. 20050558015), and the Natural Science
Foundation of Guangdong Province (No. 031541) of China. } }}
\author{Lvjun Li, Daowen Qiu\thanks{Corresponding author.  E-mail address:
issqdw@mail.sysu.edu.cn (D. Qiu)}\\ \small{Department of
Computer Science, Zhongshan University, Guangzhou 510275,}\\
\small {People's Republic of China}}
\date{ }
\maketitle

\begin{center}
\begin{minipage}{145mm}
 \vskip 2mm {\bf Abstract}
 \par
\vskip 1mm
 \hskip 5mm
In probabilistic cloning with {\it two auxiliary systems}, we
consider and compare three different protocols for the success
probabilities of cloning. We show that, in certain circumstances,
it may increase the success probability to add an auxiliary system
to the probabilistic cloning machine having {\it one auxiliary
system}, but we always can find another cloning machine with one
auxiliary system having the same success probability as that with
{\it two auxiliary systems}.


\end{minipage}
\end{center}

\begin{center}
\noindent\textbf{1. Introduction}
\end{center}

In quantum information processing, the unitarity and linearity of
quantum physics lead to some impossibilities$-$the no-cloning
theorem [1,2,3] and the no-deleting principle [4]. The linearity
of quantum theory makes an unknown quantum state unable to be
perfectly copied [1,2] and deleted [4], and two nonorthogonal
states are not allowed to be precisely cloned and deleted, as a
result of the unitarity [3,5,6], that is, for nonorthogonal pure
states $|\psi_{1}\rangle$  and $|\psi_{2}\rangle$, no physical
operation in quantum mechanic can exactly achieve the
transformation
$|\psi_{i}\rangle\rightarrow|\psi_{i}\rangle|\psi_{i}\rangle\ (i =
1, 2).$ This has been generalized to mixed states and entangled
states [7,8]. Remarkably, these restrictions provide a valuable
resource in quantum cryptography [9], because they forbid an
eavesdropper to gain information on the distributed secret key
without producing errors.

For detailed review of quantum cloning, we refer to [10]. We
briefly recall some preliminaries regarding quantum cloning. In
general, there are two kinds of cloners. One is the $universal\
quantum\ copying\ machine $ firstly introduced by Bu$\breve{z}$ek
and Hillery [11], and this kind of machines is deterministic and
does not need any information about the states to be cloned, so it
is $state-independent$. The other kind of cloners is
$state-dependent$, since it needs some information from the states
to be cloned. Furthermore, this kind of cloning machines may be
divided into three fashions of cloning: first is probabilistic
cloning proposed firstly by Duan and Guo [12,13], and then by
Chefles and Barnett [14] and Pati [15], that can clone linearly
independent states with nonzero probabilities; second is
deterministic clone first investigated by Bru\ss \ et al. [16] and
then by Chefles and Barnett [17]; third is hybrid cloners studied
by Chefles and Barnett [14], that combine deterministic cloners
with probabilistic cloning.

Recently Jozsa [18] and Horodecki {\it et al.} [19] further
clarified the no-cloning theorem and the no-deleting principle
from the viewpoint of conservation of quantum information, and in
light of
  this point of view two copies of any quantum state contain more
  information than one copy; in contrast, two classical states have
  only the same information as any one of the two states.
  Specifically, Jozsa [18] verified that if supplementary
  information, say a mixed state $\rho_{i}$ is supplemented, then there
  is a physical operation
  \begin{align}
  |\psi_{i}\rangle\otimes\rho_{i}\rightarrow|\psi_{i}\rangle|\psi_{i}\rangle
  \end{align}
  if and only if there exists physical operation
   \begin{align}
   \rho_{i}\rightarrow|\psi_{i}\rangle
   \end{align}
  where by physical operation we mean a completely positive trace-preserving map, and ${|\psi_{i}\rangle}$
  is any given finite set of pure states containing no orthogonal pair of states. This result was called {\it stronger
  no-cloning theorem} and implies that
  the supplementary information must be provided as the copy $|\psi_{i}\rangle$ itself, since the second copy can
  always be generated from the supplementary information, independently of the original copy. Therefore, this
  result may show the permanence of quantum information; that is, to get a copy of quantum state, the state
  must already exist somewhere. It is worth stressing that, if
  ${|\psi_{i}\rangle}$ contain orthogonal pairs, then the stronger
  no-cloning theorem verified by Jozsa does not hold
  again. Indeed, recently, Azuma {\it et al.} [20] proved that, for any
  pair-wise nonorthogonal set of original states
  $\{|\psi_{i}\rangle\}_{i=1,2,\cdots,n},$ if $\{|\psi_{i}\rangle\}_{i=1,2,\cdots,n}$ is irreducible, i.e.,  it
  can not be divided into two nonempty sets $S_{1}$ and $S_{2}$
  such that any state in $S_{1}$ orthogonal to any state in $S_{2}$,
  then there exists a set of supplementary states
  $\{|\phi_{i}\rangle\}_{i=1,2,\cdots,n}$ such that the following
  transformation can not be achieved over {\it local operation and
  classical communication} (LOCC):
  \begin{align}
  |\psi_{i}\rangle|\phi_{i}\rangle\ \underrightarrow{{\rm LOCC}}\
  |\psi_{i}\rangle|\psi_{i}\rangle\ (i=1,2,\cdots,n).
  \end{align}

As we stated above, cloning quantum states with a limited degree
of success has been proved always possibly. A natural issue is
that if the supplementary information is added in Duan and Guo's
probabilistic cloning [12,13] and Pati's $novel\ cloning\ machine$
(NCM) [15], then whether the optimal efficiency of the machine may
be increased. This problem was positively addressed by Azuma {\it
et al.} [21] and  Qiu [22]. Azuma {\it et al.} [21] discussed
probabilistic cloning with supplementary information contained in
the quantum states of one auxiliary system. It turns out that when
the set of input states contains only two states, the best
efficiency of producing $m$ copies is always achieved by a
two-step protocol, in which the helping party first attempts to
produce $m-1$ copies from the supplementary state, and if it
fails, then the original state is used to produce $m$ copies. When
the set of input states contains more than two states, such a
property does not hold any longer. Qiu [22] dealt with the {\it
NCM with supplementary information}, and presented an equivalent
characterization of such a quantum cloning device in terms of a
two-step cloning protocol in which the original and the
supplementary parties are only allowed to communicate with
classical channel.

In this Letter, we investigate probabilistic cloning with {\it
supplementary information contained in the quantum states of two
auxiliary systems}  (for brevity, we sometimes call it {\it two
auxiliary systems}) via three scenarios. This remainder of the
paper is organized as follows. In Section 2, we provide related
basic results and then introduce three protocols used in later
sections. We describe the three protocols in terms of different
communication channels between the original party and the two
supplementary parties: first, the original party and two
supplementary parties are in quantum communication; second, the
original party and the first supplementary party are in quantum
communication, but the first supplementary party and the second
one are in classical communication; third, the original party and
the first supplementary party are in classical communication, but
the two supplementary parties are in quantum communication.

In Section 3, we prove our main result expressed by Corollary 4,
Theorem 5 and Corollary 6; in particular, we show that, when the
two states have the same {\it a priori} probability chosen,   the
best efficiency of producing $m$ copies is achieved by the first
protocol and the third protocol. Furthermore, we also show that,
in certain circumstances, by adding an auxiliary system, we may
increase the maximum success probability of the probabilistic
cloning with supplementary information contained in the quantum
states of only {\it one} auxiliary system. However, we always can
find another probabilistic cloning with {\it one} auxiliary system
having the same success probability as that with {\it two}
auxiliary systems. Finally, in Section 4, we summarize our results
obtained, mention some potential of applications, and address a
number of related issues for further consideration.

\begin{center}
\textbf{2. Preliminaries}
\end{center}

In this section, we serve to recall Jozsa's stronger no-cloning
theorem [18] and the probabilistic cloning with supplementary
information contained in the quantum states of one auxiliary
system dealt with by Azuma {\it et al.} [21]. Then we present
three cloning protocols of probabilistic cloning with
supplementary information contained in the quantum states of two
auxiliary systems, that will be mainly discussed in Section 3.
First we recollect Duan and Guo's probabilistic cloning machine.
We denote by
\begin{align}
U_{B}(|\phi_{i}\rangle|\Sigma\rangle|P_{B}^{0}\rangle)=\sqrt{r_{i}^{B}}|\psi_{i}\rangle^{\otimes
m}|\chi_{b}\rangle|P_{B}^{i}\rangle+\sqrt{1-r_{i}^{B}}|\Psi_{bp}^{i}\rangle\
(i=1,2,...,n)
\end{align}
a machine having the following properties: (i) it receives a
quantum state from a given set $\{|\phi_{i}\rangle\}$ as an input
and return quantum states as an output $|\phi_{i}\rangle^{\otimes
m}$, together with a normalized state $|\chi_{b}\rangle$ and one
bit of classical output $|P_{B}^{i}\rangle$ indicating whether the
transformation has been successful or not; (ii) when the input
quantum state is $|\phi_{i}\rangle$, the transformation succeeds
with probability $r_{i}^{B}$, and the successful output states are
$m$ copies of $|\psi_{i}\rangle$. A necessary and sufficient
condition for the existence of Duan and Guo's probabilistic
machine is given by the following Theorem 1.
\begin{Th}[{[21]}] There exists a machine
\begin{align}
U_{B}(|\phi_{i}\rangle|\Sigma\rangle|P_{B}^{0}\rangle)=\sqrt{r_{i}^{B}}|\psi_{i}\rangle^{\otimes
m}|\chi_{b}\rangle|P_{B}^{i}\rangle+\sqrt{1-r_{i}^{B}}|\Psi_{bp}^{i}\rangle,\
i=1,2,...,n,
\end{align}
if and only if there are normalized states $|P_{B}^{0}\rangle$ and
$|P_{B}^{i}\rangle\ (i=1,2,\cdots,n)$ such that the matrix
$X-\sqrt{\Gamma}Y\sqrt{\Gamma}$ is positive semidefinite, where
$U_{B}$ is a unitary operator, $|\Sigma\rangle$ is a blank state,
$|\chi_{b}\rangle$ is a normalized state, $|\Psi_{bp}^{i}\rangle$
are normalized states of the composite system BP, and $\langle
P_{B}^{i}|\Psi_{bp}^{j}\rangle=0\ (i=1,2,\cdots,n;\
j=1,2,\cdots,n),$ $X=[\langle\phi_{i}|\phi_{j}\rangle]$, $
Y=[\langle\psi_{i}|\psi_{j}\rangle^{m}\langle
P_{B}^{i}|P_{B}^{j}\rangle]$ and $\Gamma={\rm
diag}(r_{1}^{B},r_{2}^{B},\cdots,r_{n}^{B})$ are $n \times n$
matrices.
\end{Th}

If the number of the possible original states is two, then the
necessary and sufficient condition is as follows:

\begin{Co}[{[21]}] Denote $\eta_{in}=|\langle\phi_{1}|\phi_{2}\rangle|$ and
$\eta_{out}=|\langle\psi_{1}|\psi_{2}\rangle|$. There exists a
machine
\begin{align}
U_{B}(|\phi_{i}\rangle|P_{B}^{0}\rangle)=\sqrt{r_{i}^{B}}|\psi_{i}\rangle|P_{B}^{i}\rangle+\sqrt{1-r_{i}^{B}}|\Psi_{bp}^{i}\rangle,\
i=1,2,
\end{align}
if and only if $r_{1}^{B}\geq 0, r_{2}^{B}\geq 0$, and
$\sqrt{(1-r_{1}^{B})(1-r_{2}^{B})}-\eta_{in}+\eta_{out}\sqrt{r_{1}^{B}r_{2}^{B}}\geq
0$, where $U_{B}$ is a unitary operator, $|\Psi_{bp}^{i}\rangle$
are normalized states of the composite system BP,
$|P_{B}^{i}\rangle\ (i=0,1,2)$ are probe states and normalized,
and $\langle P_{B}^{i}|\Psi_{bp}^{j}\rangle=0\ (i=1,2;\ j=1,2)$.
\end{Co}

Jozsa [18] considered how much or what kind of supplementary
information $|\phi_{i}\rangle$ is required to make two copies
$|\psi_{i}\rangle|\psi_{i}\rangle$ from the original information
$|\psi_{i}\rangle$. He showed that for any mutually nonorthogonal
set of original states $|\psi_{i}\rangle$, whenever two copies
$|\psi_{i}\rangle|\psi_{i}\rangle$ are generated with the help of
the supplementary information $|\phi_{i}\rangle$, the state
$|\psi_{i}\rangle$ can be generated from the supplementary
information $|\phi_{i}\rangle$ alone, independently of the
original state. This is described by the following theorem:

\begin{Th} [Stronger no-cloning theorem {[18]}] Let {$|\psi_{i}\rangle\
(i=1,2,...,n)$} be any finite set of pure states containing no
orthogonal pairs of states. Let $|\phi_{i}\rangle$ be any other
set of states indexed by the same labels. Then there is a physical
operation
\begin{align}
U_{AB}(|\psi_{i}\rangle|\phi_{i}\rangle|P_{AB}^{0}\rangle)=\sqrt{r_{i}^{AB}}|\psi_{i}\rangle|\psi_{i}\rangle|P_{AB}^{i}\rangle+\sqrt{1-r_{i}^{AB}}|\Psi_{abp}^{i}\rangle,\
i=1,2,\cdots,n,
\end{align}
if and only if there is a physical operation
\begin{align}
U_{B}(|\phi_{i}\rangle|P_{B}^{0}\rangle)=\sqrt{r_{i}^{B}}|\psi_{i}\rangle|P_{B}^{i}\rangle+\sqrt{1-r_{i}^{B}}|\Psi_{bp}^{i}\rangle,\
i=1,2,\cdots,n,
\end{align}
where $U_{AB}$ and $U_{B}$ are unitary operators,
$|P_{AB}^{i}\rangle\ (i=0,1,\cdots,n)$ and $|P_{B}^{i}\rangle\
(i=0,1,\cdots,n)$ are probe states and normalized,
$|\Psi_{abp}^{i}\rangle$ are normalized states of the composite
system ABP and $\langle P_{AB}^{i}|\Psi_{abp}^{j}\rangle=0\
(i=1,2,\cdots,n;\ j=1,2,\cdots,n)$, $|\Psi_{bp}^{i}\rangle$ are
normalized states of the composite system BP and $\langle
P_{B}^{i}|\Psi_{bp}^{j}\rangle=0\ (i=1,2,\cdots,n;\
j=1,2,\cdots,n)$. Here, the supplementary states
$|\phi_{i}\rangle$ $(i=1,2,...,n)$ are thought of as some pure
states.
\end{Th}

Motivated by the stronger no-cloning theorem, Azuma {\it et al.}
[21] discussed probabilistic cloning with supplementary
information contained in the quantum states of one auxiliary
system. They showed that when the set of input states contains
only two states, the best efficiency of producing $m$ copies is
always achieved by a two-step protocol, in which the helping party
first attempts to produce $m-1$ copies from the supplementary
state, and if it fails, then the original state is used to produce
$m$ copies. When the set of input states contains more than two
states, such a property does {\it not} hold any longer. Now we
consider probabilistic cloning with supplementary information
contained in the quantum states of {\it two auxiliary systems}. We
may have the following three protocols:

$Scenario\ I.$ Alice, Bob, and Victor can use two one-way quantum
channels to communicate each other. One is between Victor and Bob,
and the other is between Bob and Alice. In this case, a single
party having both the original and the supplementary information
runs a machine described by
\begin{align}
U(|\psi_{i}\rangle|\phi_{i}^{(1)}\rangle|\phi_{i}^{(2)}\rangle|\Sigma\rangle|P^{0}\rangle)=\sqrt{r_{i}^{I}}|\psi_{i}\rangle^{\otimes
m}|\chi\rangle|P^{i}\rangle+\sqrt{1-r_{i}^{I}}|\Psi_{abvp}^{i}\rangle,\
i=1,2,
\end{align}
where $r_{i}^{I}$ is the success probability of cloning
$|\psi_{i}\rangle$ with the input states
$|\psi_{i}\rangle|\phi_{i}^{(1)}\rangle|\phi_{i}^{(2)}\rangle$,
$|\Sigma\rangle$ is a blank state, $|P^{i}\rangle\ (i=0,1,2)$ are
normalized states of the probe $P$, $|\Psi_{abvp}^{i}\rangle$ are
normalized states of the composite system $ABVP$, $\langle
P^{i}|\Psi_{abvp}^{j}\rangle=0\ (i=1,2;\ j=1,2)$, $U$ is a unitary
operator, and, here and in the follow-up, $|\chi\rangle$ is a
normalized state, playing as a memory.

Before presenting Scenario II, we should understand the fact that
{\it in the same condition, the success probability of cloning
$m-1$ copies is not less than that of cloning $m$ copies ($m \geq
2$).}

According to Corollary 2, the condition required by cloning $m$
copies is that
\begin{align}
\sqrt{(1-r_{1}^{B})(1-r_{2}^{B})}-\eta_{in}+|\alpha|^{m}\sqrt{r_{1}^{B}r_{2}^{B}}\geq
0,\ i=1,2,
\end{align}
and the condition required by cloning $m-1$ copies is that
\begin{align}
\sqrt{(1-r_{1}^{B})(1-r_{2}^{B})}-\eta_{in}+|\alpha|^{m-1}\sqrt{r_{1}^{B}r_{2}^{B}}\geq
0,\ i=1,2.
\end{align}
It is clear that if $r_{i}^{B}$ satisfy Eq. (10), they also
satisfy Eq. (11). That is to say, if we succeed in cloning $m$
copies, we can clone $m-1$ copies in the same success probability.
So, the success probability of cloning $m-1$ copies is not less
than that of cloning $m$ copies.

$Scenario\ II.$ If the original information is held by Alice, and
the supplementary information by Bob and Victor, respectively. In
this case, Victor and Bob can use a one-way classical channel. Bob
and Alice can use a one-way quantum channel. In this Scenario,
there are two  ways to clone $m$ copies of input state
$|\psi_{i}\rangle$.

The first one is that: Victor, who possesses the supplementary
state $|\phi_{i}^{(2)}\rangle$, first runs the machine
\begin{align}
U_{V}(|\phi_{i}^{(2)}\rangle|\Sigma\rangle|P_{V}^{0}\rangle)=\sqrt{r_{i}^{V}}|\psi_{i}\rangle^{\otimes
m-1}|\chi_{v}\rangle|P_{V}^{i}\rangle+\sqrt{1-r_{i}^{V}}|\Psi_{vp}^{i}\rangle,\
i=1,2,
\end{align}
where $r_{i}^{V}$ is the success probability of cloning
$|\psi_{i}\rangle$ with the input state $|\phi_{i}^{(2)}\rangle$,
$|\Sigma\rangle$ is a blank state, and $|P_{V}^{i}\rangle\
(i=0,1,2)$ are normalized states of the probe $P$,
$|\Psi_{vp}^{i}\rangle$ are normalized states of the composite
system $VP$, $\langle P_{V}^{i}|\Psi_{vp}^{j}\rangle=0\ (i=1,2;\
j=1,2)$, $U_{V}$ is a unitary operator, $|\chi_{v}\rangle$ is a
normalized state, then Victor tells Alice and Bob whether the
trial was successful or not. In the successful case, Alice and Bob
just leave their states $|\psi_{i}\rangle|\phi_{i}^{(1)}\rangle$
as they are, and hence they obtain $m$ copies in total. If
Victor's attempt has failed, Alice and Bob run the machine
\begin{align}
U_{AB}(|\psi_{i}\rangle|\phi_{i}^{(1)}\rangle|\Sigma\rangle|P_{AB}^{0}\rangle)=\sqrt{r_{i}^{AB}}|\psi_{i}\rangle^{\otimes
m}|\chi_{ab}\rangle|P_{AB}^{i}\rangle+\sqrt{1-r_{i}^{AB}}|\Psi_{abp}^{i}\rangle,\
i=1,2,
\end{align}
where $r_{i}^{AB}$ is the success probability of cloning
$|\psi_{i}\rangle$ with the input states
$|\psi_{i}\rangle|\phi_{i}^{(1)}\rangle$, $|\Sigma\rangle$ is a
blank state, and $|P_{AB}^{i}\rangle\ (i=0,1,2)$ are normalized
states of the probe $P$, $|\Psi_{abp}^{i}\rangle$ are normalized
states of the composite system $ABP$, $\langle
P_{AB}^{i}|\Psi_{abp}^{j}\rangle=0\ (i=1,2;\ j=1,2)$, $U_{AB}$ is
a unitary operator, $|\chi_{ab}\rangle$ is a normalized state. So,
the total success probability of cloning $|\psi_{i}\rangle$ for
input states
$|\psi_{i}\rangle|\phi_{i}^{(1)}\rangle|\phi_{i}^{(2)}\rangle$ in
this protocol is given by
$r_{i}^{II}=r_{i}^{V}+(1-r_{i}^{V})r_{i}^{AB}.$ If we change the
form of the equality, we get that
$r_{i}^{II}=r_{i}^{AB}+(1-r_{i}^{AB})r_{i}^{V}.$

The second one is as follows: Alice and Bob first run the machine
(13), then Bob tells Victor whether the trial was successful or
not. In the successful case, Victor just leaves his state
$|\phi_{i}^{(2)}\rangle$ as it is, and hence they obtain $m$
copies in total. If Alice and Bob's attempt has failed, Victor
runs the machine
\begin{align}
U_{V}^{'}(|\phi_{i}^{(2)}\rangle|\Sigma\rangle|P_{V}^{'0}\rangle)=\sqrt{r_{i}^{'V}}|\psi_{i}\rangle^{\otimes
m}|\chi_{v}^{'}\rangle|P_{V}^{'i}\rangle+\sqrt{1-r_{i}^{'V}}|\Psi_{vp}^{'i}\rangle,\
i=1,2,
\end{align}
where $r_{i}^{'V}$ is the success probability of cloning
$|\psi_{i}\rangle$ with the input state $|\phi_{i}^{(2)}\rangle$,
$|\Sigma\rangle$ is a blank state, and $|P_{V}^{'i}\rangle\
(i=0,1,2)$ are normalized states of the probe $P$,
$|\Psi_{vp}^{'i}\rangle$ are normalized states of the composite
system $VP$, $\langle P_{V}^{'i}|\Psi_{vp}^{'j}\rangle=0\ (i=1,2;\
j=1,2)$, $U_{V}^{'}$ is a unitary operator, $|\chi_{v}^{'}\rangle$
is a normalized state. So, the total success probability of
cloning $|\psi_{i}\rangle$ for input states
$|\psi_{i}\rangle|\phi_{i}^{(1)}\rangle|\phi_{i}^{(2)}\rangle$ in
this protocol is given by
$r_{i}^{'II}=r_{i}^{AB}+(1-r_{i}^{AB})r_{i}^{'V}.$

According to the fact that the success probability of cloning
$m-1$ copies is not less than that of cloning $m$ copies, we have
$r_{i}^{V}\geq r_{i}^{'V}$. So, $r_{i}^{III}\geq r_{i}^{'III}$.
That is to say, in Scenario II, the optimal performance is always
achieved by the first way.

$Scenario\ III.$ If the original information is held by Alice, and
the supplementary information by Bob and Victor, respectively. In
this case, Victor and Bob can use a one-way quantum channel. Bob
and Alice can use a one-way classical channel. Similar to Scenario
II, in this case, we can prove that the optimal performance is
always achieved as follows: Victor and Bob, who possess the
supplementary states $|\phi_{i}^{(1)}\rangle$ and
$|\phi_{i}^{(2)}\rangle$, respectively, first run the machine
\begin{align}
U_{BV}(|\phi_{i}^{(1)}\rangle|\phi_{i}^{(2)}\rangle|\Sigma\rangle|P_{BV}^{0}\rangle)=\sqrt{r_{i}^{BV}}|\psi_{i}\rangle^{\otimes
m-1}|\chi_{bv}\rangle|P_{BV}^{i}\rangle+\sqrt{1-r_{i}^{BV}}|\Psi_{bvp}^{i}\rangle,\
i=1,2,
\end{align}
where $r_{i}^{BV}$ is the success probability of cloning
$|\psi_{i}\rangle$ with the input states $|\phi_{i}^{(1)}\rangle$
and $|\phi_{i}^{(2)}\rangle$, $|\Sigma\rangle$ is a blank state
and $|P_{BV}^{i}\rangle\ (i=0,1,2)$ are normalized states of the
probe $P$, $|\Psi_{bvp}^{i}\rangle$ are normalized states of the
composite system $BVP$, $\langle
P_{BV}^{i}|\Psi_{bvp}^{j}\rangle=0\ (i=1,2;\ j=1,2)$, $U_{BV}$ is
a unitary operator, $|\chi_{bv}\rangle$ is a normalized state.
Then Victor and Bob tell Alice whether the trial was successful or
not. In the successful case, Alice just leaves her state
$|\psi_{i}\rangle$ as it is, and hence they obtain $m$ copies in
total. If the attempt of Victor and Bob has failed, Alice runs the
machine
\begin{align}
U_{A}(|\psi_{i}\rangle|\Sigma\rangle|P_{A}^{0}\rangle)=\sqrt{r_{i}^{A}}|\psi_{i}\rangle^{\otimes
m}|\chi_{a}\rangle|P_{A}^{i}\rangle+\sqrt{1-r_{i}^{A}}|\Psi_{ap}^{i}\rangle,\
i=1,2,
\end{align}
where $r_{i}^{A}$ is the success probability of cloning
$|\psi_{i}\rangle$ with the input state $|\psi_{i}\rangle$,
$|\Sigma\rangle$ is a blank state, and $|P_{A}^{i}\rangle\
(i=0,1,2)$ are normalized states of the probe $P$,
$|\Psi_{ap}^{i}\rangle$ are normalized states of the composite
system $AP$, $\langle P_{A}^{i}|\Psi_{ap}^{j}\rangle=0\ (i=1,2;\
j=1,2)$, $U_{A}$ is a unitary operator, $|\chi_{a}\rangle$ is a
normalized state. So, the total success probability of cloning
$|\psi_{i}\rangle$ for input states
$|\psi_{i}\rangle|\phi_{i}^{(1)}\rangle|\phi_{i}^{(2)}\rangle$ in
this protocol is given by
$r_{i}^{III}=r_{i}^{BV}+(1-r_{i}^{BV})r_{i}^{A}$.

Now, based on the three protocols above, we are ready to calculate
their maximum success probabilities and investigate their
relationships in next section.

\begin{center}
\textbf{3. Probabilistic cloning with supplementary information
contained in the quantum states of two auxiliary systems }
\end{center}

Throughout this paper, we consider $\{|\psi_{1}\rangle,\
|\psi_{2}\rangle\}$ as the original states.
$\{|\phi_{1}^{(1)}\rangle,\ |\phi_{2}^{(1)}\rangle\}$ and
$\{|\phi_{1}^{(2)}\rangle, |\phi_{2}^{(2)}\rangle\}$ are
supplementary information contained in the quantum states of two
auxiliary systems. For convenience, we denote
$\langle\psi_{1}|\psi_{2}\rangle=\alpha,\
\langle\phi_{1}^{(1)}|\phi_{2}^{(1)}\rangle=\beta,\
\langle\phi_{1}^{(2)}|\phi_{2}^{(2)}\rangle=\gamma.$ Let $P_{i}$
be the prior probability of $|\psi_{i}\rangle$,  and, let $r^{I},\
r^{II},\ r^{III}$ be the total success probabilities in $Scenario\
I,\ II,\ III$, respectively. Let $r_{i}^{I},\ r_{i}^{II},\
r_{i}^{III}$ denote the success probabilities of cloning
$|\psi_{i}\rangle$ in $Scenario\ I,\ II,\ III$, respectively.
Clearly, we have the following relationships:
$r^{I}=P_{1}r_{1}^{I}+P_{2}r_{2}^{I}$,
$r^{II}=P_{1}r_{1}^{II}+P_{2}r_{2}^{II}$,
$r^{III}=P_{1}r_{1}^{III}+P_{2}r_{2}^{III}$. Moreover, by
$r_{max}^{I},\ r_{max}^{II},\ r_{max}^{III}$ we mean the maximum
values of $r^{I},\ r^{II},\ r^{III}$, respectively.

Before presenting the main result, we verify a corollary as
follows, which will be used in the proof of  Theorem 5.

\begin{Co}
Denote $\eta_{in}=|\langle\phi_{1}|\phi_{2}\rangle|$ and
$\eta_{out}=|\langle\psi_{1}|\psi_{2}\rangle|$. Suppose that
$\eta_{in}>\eta_{out}$. If there exists a machine
\begin{align}
U_{B}(|\phi_{i}\rangle|P_{B}^{0}\rangle)=\sqrt{r_{i}^{B}}|\psi_{i}\rangle|P_{B}^{i}\rangle+\sqrt{1-r_{i}^{B}}|\Psi_{bp}^{i}\rangle,\
i=1,2,
\end{align}
where $U_{B}$ is a unitary operator, $|\Psi_{bp}^{i}\rangle$ are
normalized states of the composite system BP, $|P_{B}^{i}\rangle\
(i=0,1,2)$ are probe states and normalized, $\langle
P_{B}^{i}|\Psi_{bp}^{j}\rangle=0\ (i=1,2;\ j=1,2)$, $r_{i}^{B}$
(i=1,2) are the success probabilities of cloning
$|\psi_{i}\rangle$, then we can conclude that $r_{1}^{B}$ and
$r_{2}^{B}$ satisfy
\begin{align}
\frac{r_{1}^{B}+r_{2}^{B}}{2}\leq\frac{1-\eta_{in}}{1-\eta_{out}}.
\end{align}
\end{Co}

\noindent \textbf{Proof.} According to Corollary 2, if there
exists a machine
\begin{align*}
U_{B}(|\phi_{i}\rangle|P_{B}^{0}\rangle)=\sqrt{r_{i}^{B}}|\psi_{i}\rangle|P_{B}^{i}\rangle+\sqrt{1-r_{i}^{B}}|\Psi_{bp}^{i}\rangle,\
i=1,2,
\end{align*}
where $U_{B}$ is a unitary operator, $|\Psi_{bp}^{i}\rangle$ are
normalized states of the composite system BP, $|P_{B}^{i}\rangle\
(i=0,1,2)$ are probe states and normalized, $\langle
P_{B}^{i}|\Psi_{bp}^{j}\rangle=0\ (i=1,2;\ j=1,2)$, then we have
\begin{align}
\sqrt{(1-r_{1}^{B})(1-r_{2}^{B})}-\eta_{in}+\eta_{out}\sqrt{r_{1}^{B}r_{2}^{B}}\geq
0.
\end{align}
And it is known that
\begin{eqnarray}
\begin{split}
&\sqrt{(1-r_{1}^{B})(1-r_{2}^{B})}-\eta_{in}+\eta_{out}\sqrt{r_{1}^{B}r_{2}^{B}}
\\\leq&\frac{(1-r_{1}^{B})+(1-r_{2}^{B})}{2}-\eta_{in}+\eta_{out}\frac{r_{1}^{B}+r_{2}^{B}}{2}
\end{split}
\end{eqnarray}
where the equality holds if and only if $r_{1}^{B}=r_{2}^{B}$.
Using Eqs. (19) and (20), we get
\begin{align}
\frac{(1-r_{1}^{B})+(1-r_{2}^{B})}{2}-\eta_{in}+\eta_{out}\frac{r_{1}^{B}+r_{2}^{B}}{2}\geq0.
\end{align}
That is to say,
\begin{align}
1-\frac{r_{1}^{B}+r_{2}^{B}}{2}-\eta_{in}+\eta_{out}\frac{r_{1}^{B}+r_{2}^{B}}{2}\geq0.
\end{align}
Thus, we conclude that
\begin{align}
\frac{r_{1}^{B}+r_{2}^{B}}{2}\leq\frac{1-\eta_{in}}{1-\eta_{out}},
\end{align}
where the equality holds if and only if $r_{1}^{B}=r_{2}^{B}$. \qed\\

Now we can present the main result.

\begin{Th}
The relationships of the three Scenarios above are as follows:

1. When $|\beta|\leq|\alpha|^{m-1}$ or
$|\gamma|\leq|\alpha|^{m-1}$, the maximum success probabilities of
the three Scenarios are equal to 1.

2. When $|\beta|>|\alpha|^{m-1}$ and $|\gamma|>|\alpha|^{m-1}$,
there are two cases:

(1) If $|\alpha|^{2m-2}<|\beta\gamma|\leq|\alpha|^{m-1}$, we have
$r_{max}^{I}=r_{max}^{III}> r_{max}^{II}$.

(2) If $|\alpha|^{m-1}<|\beta\gamma|\leq1$ and $P_{1}=P_{2}$, we
conclude that $r_{max}^{I}=r_{max}^{III}\geq r_{max}^{II}$, and
only if $|\gamma|=1$ or $|\beta|=1$, we get the equality
$r_{max}^{I}=r_{max}^{III}= r_{max}^{II}$.

\end{Th}

\noindent \textbf{Proof.} 1. When $|\gamma|\leq |\alpha|^{m-1}$ or
$|\beta|\leq|\alpha|^{m-1}$, from Corollary 2, there exists a
machine described by
\begin{align}
U_{V}(|\phi_{i}^{(2)}\rangle|\Sigma\rangle|P_{V}^{0}\rangle)=\sqrt{r_{i}^{V}}|\psi_{i}\rangle^{\otimes
m-1}|\chi_{v}\rangle|P_{V}^{i}\rangle+\sqrt{1-r_{i}^{V}}|\Psi_{vp}^{i}\rangle,
\ i=1,2,
\end{align}
satisfying $r_{1}^{V}=r_{2}^{V}=1$, where $r_{i}^{V}$ is the
success probability with the input state $|\phi_{i}^{(2)}\rangle$,
$|\Sigma\rangle$ is a blank state, and $|P_{V}^{i}\rangle\
(i=0,1,2)$ are normalized states of the probe $P$,
$|\Psi_{vp}^{i}\rangle$ are normalized states of the composite
system $VP$, $\langle P_{V}^{i}|\Psi_{vp}^{j}\rangle=0\ (i=1,2;\
j=1,2)$, $U_{V}$ is a unitary operator, $|\chi_{v}\rangle$ is a
normalized state. So, we have the equality
$r_{i}^{II}=r_{i}^{V}+(1-r_{i}^{V})r_{i}^{AB}=1$. Or, there exists
a machine
\begin{align}
U_{AB}(|\psi_{i}\rangle|\phi_{i}^{(1)}\rangle|\Sigma\rangle|P_{AB}^{0}\rangle)=\sqrt{r_{i}^{AB}}|\psi_{i}\rangle^{\otimes
m}|\chi_{ab}\rangle|P_{AB}^{i}\rangle+\sqrt{1-r_{i}^{AB}}|\Psi_{abp}^{i}\rangle,\
i=1,2,
 \end{align}
satisfying $r_{1}^{AB}=r_{2}^{AB}=1$, where $r_{i}^{AB}$ is the
success probability with the input states
$|\psi_{i}\rangle|\phi_{i}^{(1)}\rangle$, $|\Sigma\rangle$ is a
blank state, and $|P_{AB}^{i}\rangle\ (i=0,1,2)$ are normalized
states of the probe $P$, $|\psi_{abp}^{i}\rangle$ are normalized
states of the composite system $ABP$, $\langle
P_{AB}^{i}|\Psi_{abp}^{j}\rangle=0\ (i=1,2;\ j=1,2)$, $U_{AB}$ is
a unitary operator, $|\chi_{ab}\rangle$ is a normalized state. We
also have $r_{i}^{II}=r_{i}^{V}+(1-r_{i}^{V})r_{i}^{AB}=1$.

Because $|\gamma|\leq|\alpha|^{m-1}$ or
$|\beta|\leq|\alpha|^{m-1}$, we have
$|\alpha\beta\gamma|\leq|\alpha|^{m-1}$, and thus there exists a
machine
\begin{align}
U(|\psi_{i}\rangle|\phi_{i}^{(1)}\rangle|\phi_{i}^{(2)}\rangle|\Sigma\rangle|P^{0}\rangle)=\sqrt{r_{i}^{I}}|\psi_{i}\rangle^{\otimes
m}|\chi\rangle|P^{i}\rangle+\sqrt{1-r_{i}^{I}}|\Psi_{abvp}^{i}\rangle,\
i=1,2,
\end{align}
satisfying $r_{1}^{I}=r_{2}^{I}=1$, where $r_{i}^{I}$ is the
success probability with the input states
$|\psi_{i}\rangle|\phi_{i}^{(1)}\rangle|\phi_{i}^{(2)}\rangle$,
$|\Sigma\rangle$ is a blank state, and $|P^{i}\rangle\ (i=0,1,2)$
are normalized states of the probe $P$, $|\Psi_{abvp}^{i}\rangle$
are normalized states of the composite system $ABVP$, $\langle
P^{i}|\Psi_{abvp}^{j}\rangle=0\ (i=1,2;\ j=1,2)$, $U$ is a unitary
operator, $|\chi\rangle$ is a normalized state.

Because $|\gamma|\leq|\alpha|^{m-1}$ or
$|\beta|\leq|\alpha|^{m-1}$, we have
$|\beta\gamma|\leq|\alpha|^{m}$, and thus there exists a machine
\begin{align}
U_{BV}(|\phi_{i}^{(1)}\rangle|\phi_{i}^{(2)}\rangle|\Sigma\rangle|P_{BV}^{0}\rangle)=\sqrt{r_{i}^{BV}}|\psi_{i}\rangle^{\otimes
m-1}|\chi_{bv}\rangle|P_{BV}^{i}\rangle+\sqrt{1-r_{i}^{BV}}|\Psi_{bvp}^{i}\rangle,\
i=1,2,
\end{align}
satisfying $r_{1}^{BV}=r_{2}^{BV}=1$, where $r_{i}^{BV}$ is the
success probability with the input states $|\phi_{i}^{(1)}\rangle$
and $|\phi_{i}^{(2)}\rangle$, $|\Sigma\rangle$ is a blank state
and $|P_{BV}^{i}\rangle\ (i=0,1,2)$ are normalized states of the
probe $P$, $|\Psi_{bvp}^{i}\rangle$ are normalized states of the
composite system $BVP$, $\langle
P_{BV}^{i}|\Psi_{bvp}^{j}\rangle=0\ (i=1,2;\ j=1,2)$, $U_{BV}$ is
a unitary operator, $|\chi_{bv}\rangle$ is a normalized state. So,
we have $r_{i}^{III}=r_{i}^{BV}+(1-r_{i}^{BV})r_{i}^{A}=1$.

As a result, we conclude that
$r_{i}^{III}=r_{i}^{II}=r_{i}^{I}=1$. So, in this case,
$r_{max}^{I}=r_{max}^{III}= r_{max}^{II}=1$.

2.  When $|\beta|>|\alpha|^{m-1}$ and $|\gamma|>|\alpha|^{m-1}$,
we have the two cases as follows:

1) If $|\alpha|^{2m-2}<|\beta\gamma|\leq|\alpha|^{m-1}$, from
Corollary 2, there exists a machine
\begin{align}
U(|\psi_{i}\rangle|\phi_{i}^{(1)}\rangle|\phi_{i}^{(2)}\rangle|\Sigma\rangle|P^{0}\rangle)=\sqrt{r_{i}^{I}}|\psi_{i}\rangle^{\otimes
m}|\chi\rangle|P^{i}\rangle+\sqrt{1-r_{i}^{I}}|\Psi_{abvp}^{i}\rangle,\
i=1,2,
\end{align}
with $r_{1}^{I}=r_{2}^{I}=1$, where $r_{i}^{I}$ is the success
probability with the input states
$|\psi_{i}\rangle|\phi_{i}^{(1)}\rangle|\phi_{i}^{(2)}\rangle$,
$|\Sigma\rangle$ is a blank state, and $|P^{i}\rangle\ (i=0,1,2)$
are normalized states of the probe $P$, $|\Psi_{abvp}^{i}\rangle$
are normalized states of the composite system $ABVP$, $\langle
P^{i}|\Psi_{abvp}^{j}\rangle=0\ (i=1,2;\ j=1,2)$, $U$ is a unitary
operator, $|\chi\rangle$ is a normalized state. So, we conclude
\begin{align}r_{max}^{I}=1.\end{align}
And there also exists a machine
\begin{align}
U_{BV}(|\phi_{i}^{(1)}\rangle|\phi_{i}^{(2)}\rangle|\Sigma\rangle|P_{BV}^{0}\rangle)=\sqrt{r_{i}^{BV}}|\psi_{i}\rangle^{\otimes
m-1}|\chi_{bv}\rangle|P_{BV}^{i}\rangle+\sqrt{1-r_{i}^{BV}}|\Psi_{bvp}^{i}\rangle,\
i=1,2,
\end{align}
with $r_{1}^{BV}=r_{2}^{BV}=1$, where $r_{i}^{BV}$ is the success
probability with the input states $|\phi_{i}^{(1)}\rangle$ and
$|\phi_{i}^{(2)}\rangle$, $|\Sigma\rangle$ is a blank state and
$|P_{BV}^{i}\rangle\ (i=0,1,2)$ are normalized states of the probe
$P$, $|\Psi_{bvp}^{i}\rangle$ are normalized states of the
composite system $BVP$, $\langle
P_{BV}^{i}|\Psi_{bvp}^{j}\rangle=0\ (i=1,2;\ j=1,2)$, $U_{BV}$ is
a unitary operator, $|\chi_{bv}\rangle$ is a normalized state. So,
we get $r_{i}^{III}=r_{i}^{BV}+(1-r_{i}^{BV})r_{i}^{A}=1$, and
$r^{III}=P_{1}r_{1}^{III}+p_{2}r_{2}^{III}=1$. As a consequence,
we have
\begin{align}r_{max}^{III}=1.\end{align}
Since $r_{i}^{II}\leq1$ holds in any case and the success
probability of $Scenario\ II$ can not be 1 in this case, together
with $r^{II}=P_{1}r_{1}^{II}+P_{2}r_{2}^{II}$,  we  clearly have
$r^{II}<1$. So, we conclude
\begin{align}
r_{max}^{II}<1.
\end{align}
Using Eqs. (29,31,32), we conclude that
$r_{max}^{I}=r_{max}^{III}=1> r_{max}^{II}$.

2) If $|\alpha|^{m-1}<|\beta\gamma|\leq1$ and
$P_{1}=P_{2}=\frac{1}{2}$, we investigate the relationships
between the three Scenarios as follows:

In $Scenario\ I$, there exists a machine
\begin{align}
U(|\psi_{i}\rangle|\phi_{i}^{(1)}\rangle|\phi_{i}^{(2)}\rangle|\Sigma\rangle|P^{0}\rangle)=\sqrt{r_{i}^{I}}|\psi_{i}\rangle^{\otimes
m}|\chi\rangle|P^{i}\rangle+\sqrt{1-r_{i}^{I}}|\Psi_{abvp}^{i}\rangle\,
i=1,2,
\end{align}
where $r_{i}^{I}$ is the success probability with the input states
$|\psi_{i}\rangle|\phi_{i}^{(1)}\rangle|\phi_{i}^{(2)}\rangle$,
$|\Sigma\rangle$ is a blank state, and $|P^{i}\rangle\ (i=0,1,2)$
are normalized states of the probe $P$, $|\Psi_{abvp}^{i}\rangle$
are normalized states of the composite system $ABVP$, $\langle
P^{i}|\Psi_{abvp}^{j}\rangle=0\ (i=1,2;\ j=1,2)$, $U$ is a unitary
operator, $|\chi\rangle$ is a normalized state. In this machine,
$\eta_{in}=|\alpha\beta\gamma|$ and $\eta_{out}=|\alpha|^{m}$.
According to $Corollary\ 4$, we conclude that
\begin{align}
r^{I}=\frac{r_{1}^{I}+r_{2}^{I}}{2}\leq\frac{1-|\alpha\beta\gamma|}{1-|\alpha|^{m}}.
\end{align}
The equality (34) holds if and only if $r_{1}^{I}=r_{2}^{I}$.
Therefore, we have
\begin{align}
r_{max}^{I}=\frac{1-|\alpha\beta\gamma|}{1-|\alpha|^{m}}.
\end{align}

In $Scenario\ II$, there also exists a machine
\begin{align}
U_{V}(|\phi_{i}^{(2)}\rangle|\Sigma\rangle|P_{V}^{0}\rangle)=\sqrt{r_{i}^{V}}|\psi_{i}\rangle^{\otimes
m-1}|\chi_{v}\rangle|P_{V}^{i}\rangle+\sqrt{1-r_{i}^{V}}|\Psi_{vp}^{i}\rangle,\
i=1,2,
\end{align}
where $r_{i}^{V}$ is the success probability with the input state
$|\phi_{i}^{(2)}\rangle$, $|\Sigma\rangle$ is a blank state, and
$|P_{V}^{i}\rangle\ (i=0,1,2)$ are normalized states of the probe
$P$, $|\Psi_{vp}^{i}\rangle$ are normalized states of the
composite system $VP$, $\langle P_{V}^{i}|\Psi_{vp}^{j}\rangle=0\
(i=1,2;\ j=1,2)$, $U_{V}$ is a unitary operator,
$|\chi_{v}\rangle$ is a normalized state. In this case,
$\eta_{in}=|\gamma|$ and $\eta_{out}=|\alpha|^{m}$. According to
$Corollary\ 4$, we conclude that
 \begin{align}
\frac{r_{1}^{\upsilon}+r_{2}^{\upsilon}}{2}\leq\frac{1-|\gamma|}{1-|\alpha|^{m}}.
\end{align}
And there exists another machine
\begin{align}
U_{AB}(|\psi_{i}\rangle|\phi_{i}^{(1)}\rangle|\Sigma\rangle|P_{AB}^{0}\rangle)=\sqrt{r_{i}^{AB}}|\psi_{i}\rangle^{\otimes
m}|\chi_{ab}\rangle|P_{AB}^{i}\rangle+\sqrt{1-r_{i}^{AB}}|\Psi_{abp}^{i}\rangle,\
i=1,2,
\end{align}
where $r_{i}^{AB}$ is the success probability with the input
states $|\psi_{i}\rangle|\phi_{i}^{(1)}\rangle$, $|\Sigma\rangle$
is a blank state, and $|P_{AB}^{i}\rangle\ (i=0,1,2)$ are
normalized states of the probe $P$, $|\psi_{abp}^{i}\rangle$ are
normalized states of the composite system $ABP$, $\langle
P_{AB}^{i}|\Psi_{abp}^{j}\rangle=0\ (i=1,2;\ j=1,2)$, $U_{AB}$ is
a unitary operator, $|\chi_{ab}\rangle$ is a normalized state. In
this machine, $\eta_{in}=|\alpha\beta|$ and
$\eta_{out}=|\alpha|^{m}$. According to $Corollary\ 4$, we
conclude that
\begin{align}
\frac{r_{1}^{AB}+r_{2}^{AB}}{2}\leq\frac{1-|\alpha\beta|}{1-|\alpha|^{m}}.
\end{align}
So, we have
\begin{eqnarray}
r^{II}&=&\frac{r_{1}^{V}+(1-r_{1}^{V})r_{1}^{AB}}{2}+\frac{r_{2}^{V}+(1-r_{2}^{V})r_{2}^{AB}}{2}
\\&=&\frac{r_{1}^{V}+r_{2}^{V}}{2}+\frac{r_{1}^{AB}+r_{2}^{AB}}{2}-\frac{r_{1}^{V}r_{1}^{AB}+r_{2}^{V}r_{2}^{AB}}{2}
\\&\leq&\frac{1-|\gamma|}{1-|\alpha|^{m-1}}+\frac{1-|\alpha\beta|}{1-|\alpha|^{m}}-\frac{1-|\gamma|}{1-|\alpha|^{m-1}}\frac{1-|\alpha\beta|}{1-|\alpha|^{m}}
\\&=&\frac{(1-|\gamma|)(|\alpha\beta|-|\alpha|^{m})+(1-|\alpha\beta|)(1-|\alpha|^{m-1})}{(1-|\alpha|^{m})(1-|\alpha|^{m-1})}.
\end{eqnarray}
Also, the above equality holds if and only if
$r_{1}^{AB}=r_{2}^{AB}$ and $r_{1}^{V}=r_{2}^{V}$. Consequently,
we have
\begin{align}
r_{max}^{II}=\frac{(1-|\gamma|)(|\alpha\beta|-|\alpha|^{m})+(1-|\alpha\beta|)(1-|\alpha|^{m-1})}{(1-|\alpha|^{m})(1-|\alpha|^{m-1})}.
\end{align}

As well, we can compare the success probability of Scenario I with
that of Scenario II. Because
\begin{eqnarray}
r_{max}^{II}-r_{max}^{I}&=&\frac{(1-|\gamma|)(|\alpha\beta|-|\alpha|^{m})+(1-|\alpha\beta|)(1-|\alpha|^{m-1})}{(1-|\alpha|^{m})(1-|\alpha|^{m-1})}-\frac{1-|\alpha\beta\gamma|}{1-|\alpha|^{m}}
\\&=&\frac{(1-|\gamma|)|\alpha|^m(|\beta|-1)}{(1-|\alpha|^{m})(1-|\alpha|^{m-1})}
\\&\leq&0,
\end{eqnarray}
we get \begin{align}r_{max}^{II}\leq r_{max}^{I}.\end{align}

In $scenario\ III$, there exists a machine
\begin{align}
U_{BV}(|\phi_{i}^{(1)}\rangle|\phi_{i}^{(2)}\rangle|\Sigma\rangle|P_{BV}^{0}\rangle)=\sqrt{r_{i}^{BV}}|\psi_{i}\rangle^{\otimes
m-1}|\chi_{bv}\rangle|P_{BV}^{i}\rangle+\sqrt{1-r_{i}^{BV}}|\Psi_{bvp}^{i}\rangle,\
i=1,2,
\end{align}
where $r_{i}^{BV}$ is the success probability with the input
states $|\phi_{i}^{(1)}\rangle$ and $|\phi_{i}^{(2)}\rangle$,
$|\Sigma\rangle$ is a blank state and $|P_{BV}^{i}\rangle\
(i=0,1,2)$ are normalized states of the probe $P$,
$|\Psi_{bvp}^{i}\rangle$ are normalized states of the composite
system $BVP$, $\langle P_{BV}^{i}|\Psi_{bvp}^{j}\rangle=0\
(i=1,2;\ j=1,2)$, $U_{BV}$ is a unitary operator,
$|\chi_{bv}\rangle$ is a normalized state. In this machine,
$\eta_{in}=|\beta\gamma|$ and $\eta_{out}=|\alpha|^{m-1}$.
According to $Corollary\ 4$, we conclude that
\begin{align}
\frac{r_{1}^{BV}+r_{2}^{BV}}{2}\leq\frac{1-|\beta\gamma|}{1-|\alpha|^{m-1}}.
\end{align}
And there exists a machine

\begin{align}
U_{A}(|\psi_{i}\rangle|\Sigma\rangle|P_{A}^{0}\rangle)=\sqrt{r_{i}^{A}}|\psi_{i}\rangle^{\otimes
m}|\chi_{a}\rangle|P_{A}^{i}\rangle+\sqrt{1-r_{i}^{A}}|\Psi_{ap}^{i}\rangle,\
i=1,2,
\end{align}
where $r_{i}^{A}$ is the success probability with the input state
$|\psi_{i}\rangle$, $|\Sigma\rangle$ is a blank state, and
$|P_{A}^{i}\rangle\ (i=0,1,2)$ are normalized states of the probe
$P$, $|\Psi_{ap}^{i}\rangle$ are normalized states of the
composite system $AP$, $\langle P_{A}^{i}|\Psi_{ap}^{j}\rangle=0\
(i=1,2;\ j=1,2)$, $U_{A}$ is a unitary operator,
$|\chi_{a}\rangle$ is a normalized state. In this case,
$\eta_{in}=|\alpha|$ and $\eta_{out}=|\alpha|^{m}$. According to
$Corollary\ 4$, we conclude that
\begin{align}
\frac{r_{1}^{A}+r_{2}^{A}}{2}\leq\frac{1-|\alpha|}{1-|\alpha|^{m}}.
\end{align}
So, we have
\begin{eqnarray}
r^{III}&=&\frac{r_{1}^{BV}+(1-r_{1}^{BV})r_{1}^{A}}{2}+\frac{r_{2}^{BV}+(1-r_{2}^{BV})r_{2}^{A}}{2}
\\&=&\frac{r_{1}^{BV}+r_{2}^{BV}}{2}+\frac{r_{1}^{A}+r_{2}^{A}}{2}-\frac{r_{1}^{BV}r_{1}^{A}+r_{2}^{BV}r_{2}^{A}}{2}
\\&\leq&\frac{1-|\beta\gamma|}{1-|\alpha|^{m-1}}+\frac{1-|\alpha|}{1-|\alpha|^{m}}-\frac{1-|\beta\gamma|}{1-|\alpha|^{m-1}}\frac{1-|\alpha|}{1-|\alpha|^{m}}
\\&=&\frac{(1-|\beta\gamma|)(1-|\alpha|^{m})+(1-|\alpha|)(1-|\alpha|^{m-1})-(1-|\beta\gamma|)(1-|\alpha|)}{(1-|\alpha|^{m})(1-|\alpha|^{m-1})}
\\&=&\frac{1-|\alpha\beta\gamma|}{1-|\alpha|^{m}}.
\end{eqnarray}
Also, the above equality holds if and only if
$r_{1}^{AB}=r_{2}^{AB}$ and $r_{1}^{V}=r_{2}^{V}$. As a result,
\begin{align}
r_{max}^{III}=\frac{1-|\alpha\beta\gamma|}{1-|\alpha|^{m}}.
\end{align}
As well, we can compare the success probability of $Scenario\ I$
with that of $Scenario\ III$. Because
\begin{align}
r_{max}^{III}-r_{max}^{I}=\frac{1-|\alpha\beta\gamma|}{1-|\alpha|^{m}}-\frac{1-|\alpha\beta\gamma|}{1-|\alpha|^{m}}=0,
\end{align}
we get \begin{align}r_{max}^{III}=r_{max}^{I}.\end{align}

By using Eqs. (45-48) and Eq. (60), we conclude that
$r_{max}^{I}=r_{max}^{III}\geq r_{max}^{II}$. If $|\gamma|\neq1$
and $|\beta|\neq1$, we have $r_{max}^{I}> r_{max}^{II}$, and only
if $|\gamma|=1$ or $|\beta|=1$, we get the equality
$r_{max}^{I}=r_{max}^{III}= r_{max}^{II}$. \qed\\

Based on the above proof, we find that, by adding one auxiliary
system, it is possible to increase the success probability for
probabilistic cloning machine with {\it one} auxiliary system.
However, given a probabilistic cloning machine with {\it two}
auxiliary systems, we always can establish an probabilistic
cloning machine with {\it one} auxiliary system, whose success
probability is the same as that with {\it two} auxiliary systems.
We further describe this by the following corollary.

\begin{Co}
When $P_{1}=P_{2},\ |\beta|>|\alpha|^{m-1},\
|\gamma|>|\alpha|^{m-1}\ and\ |\beta|\neq 1,\ |\gamma|\neq 1$, by
means of adding an auxiliary system described by states
$\{|\phi_{1}^{(2)}\rangle,\ |\phi_{2}^{(2)}\rangle\}$, we may
increase the maximum success probability of the cloning machine
with only one auxiliary system described by states
$\{|\phi_{1}^{(1)}\rangle,\ |\phi_{2}^{(1)}\rangle\}$. However,
given a probabilistic cloning machine with {\it two} auxiliary
systems described by states $\{|\phi_{1}^{(1)}\rangle,\
|\phi_{2}^{(1)}\rangle\}$ and $\{|\phi_{1}^{(2)}\rangle,\
|\phi_{2}^{(2)}\rangle\}$, respectively, we always can find
another probabilistic cloning machine with only {\it one}
auxiliary system described by states $\{|\phi_{1}^{(3)}\rangle,\
|\phi_{2}^{(3)}\rangle\}$, satisfying the condition:
$\langle\phi_{1}^{(3)}|\phi_{2}^{(3)}\rangle=\langle\phi_{1}^{(1)}|\phi_{2}^{(1)}\rangle\langle\phi_{1}^{(2)}|\phi_{2}^{(2)}\rangle$,
such that the two cloning machines have the same maximum success
probability.
\end{Co}

\noindent \textbf{Proof.} Clearly, we have
\begin{align}
r_{i}^{AB}+(1-r_{i}^{AB})r_{i}^{V}\geq r_{i}^{AB},
\end{align}
and according to $Scenario\ II$, we have
\begin{align}
r_{i}^{II}=r_{i}^{AB}+(1-r_{i}^{AB})r_{i}^{V}
\end{align}
where $r_{i}^{AB}$ are the success probabilities of cloning state
$|\psi_{i}\rangle$ with supplementary information contained in
quantum states of one auxiliary system described by
$\{|\phi_{1}^{(1)}\rangle,|\phi_{2}^{(1)}\rangle\}$. Let $r^{AB}$
be the total success probability with supplementary information
contained in the quantum states of one auxiliary system, that is
to say,
\begin{align}
r^{AB}=P_{1}r_{1}^{AB}+P_{2}r_{2}^{AB}.
\end{align}
Because $r^{II}=P_{1}r_{1}^{II}+P_{2}r_{2}^{II}$, we  conclude
that
\begin{align}
r^{II}\geq r^{AB}.
\end{align}
 So, we have \begin{align}r_{max}^{II}\geq r_{max}^{AB},\end{align}
where $r_{max}^{AB}$ denotes the maximum
value of $r^{AB}$.

According to {\it Theorem 5} and Eq. (65), if $P_{1}=P_{2},\
|\beta|>|\alpha|^{m-1},\ |\gamma|>|\alpha|^{m-1}\ and\ |\beta|\neq
1,\ |\gamma|\neq 1$, we have
\begin{align}
r_{max}^{I}> r_{max}^{AB}.
\end{align}
That is to say, we may increase the maximum success probability of
cloning through adding an auxiliary system described by states
$\{|\phi_{1}^{(2)}\rangle,\ |\phi_{2}^{(2)}\rangle\}$.

When auxiliary system described by states
$\{|\phi_{1}^{(3)}\rangle,\ |\phi_{2}^{(3)}\rangle\}$ satisfies
that
\begin{align}
\langle\phi_{1}^{(3)}|\phi_{2}^{(3)}\rangle=\langle\phi_{1}^{(1)}|\phi_{2}^{(1)}\rangle\langle\phi_{1}^{(2)}|\phi_{2}^{(2)}\rangle,
\end{align}
we can consider in $Scenario\ III$ the product state
$|\phi^{(1)}_{i}\rangle|\phi^{(2)}_{i}\rangle$ as a state
$|\phi^{(3)}_{i}\rangle$ of one auxiliary system. So, by changing
$Scenario\ III$, we can describe it as follows:

If the original information is held by Alice, and the
supplementary information $\{|\phi_{1}^{(3)}\rangle,\
|\phi_{2}^{(3)}\rangle\}$ by John. John and Alice can use a
one-way classical channel. In this case, the optimal performance
is always achieved as follows: John first runs the machine
\begin{align}
U_{J}(|\phi_{i}^{(3)}\rangle|\Sigma\rangle|P_{J}^{0}\rangle)=\sqrt{r_{i}^{J}}|\psi_{i}\rangle^{\otimes
m-1}|\chi_{j}\rangle|P_{J}^{i}\rangle+\sqrt{1-r_{i}^{J}}|\Psi_{jp}^{i}\rangle,\
i=1,2,
\end{align}
where $r_{i}^{J}$ is the success probability of cloning
$|\psi_{i}\rangle$ with the input state $|\phi_{i}^{(3)}\rangle$,
$|\Sigma\rangle$ is a blank state and $|P_{J}^{i}\rangle\
(i=0,1,2)$ are normalized states of the probe $P$,
$|\Psi_{jp}^{i}\rangle$ are normalized states of the composite
system $JP$, $\langle P_{J}^{i}|\Psi_{jp}^{j}\rangle=0\ (i=1,2;\
j=1,2)$, $U_{J}$ is a unitary operator, $|\chi_{j}\rangle$ is a
normalized state, and afterwards John tells Alice whether the
trial was successful or not. In the successful case, Alice just
leaves her state $|\psi_{i}\rangle$ as it is, and hence they
obtain $m$ copies in total. If the attempt of Victor and Bob has
failed, Alice runs the machine
\begin{align}
U_{A}(|\psi_{i}\rangle|\Sigma\rangle|P_{A}^{0}\rangle)=\sqrt{r_{i}^{A}}|\psi_{i}\rangle^{\otimes
m}|\chi_{a}\rangle|P_{A}^{i}\rangle+\sqrt{1-r_{i}^{A}}|\Psi_{ap}^{i}\rangle,\
i=1,2,
\end{align}
where $r_{i}^{A}$ is the success probability of cloning
$|\psi_{i}\rangle$ with the input state $|\psi_{i}\rangle$,
$|\Sigma\rangle$ is a blank state, and $|P_{A}^{i}\rangle\
(i=0,1,2)$ are normalized states of the probe $P$,
$|\Psi_{ap}^{i}\rangle$ are normalized states of the composite
system $AP$, $\langle P_{A}^{i}|\Psi_{ap}^{j}\rangle=0\ (i=1,2;\
j=1,2)$, $U_{A}$ is a unitary operator, $|\chi_{a}\rangle$ is a
normalized state. So, the total success probability of cloning
$|\psi_{i}\rangle$ for input state
$|\psi_{i}\rangle|\phi_{i}^{(3)}\rangle$ in this protocol is given
by $r_{i}^{III}=r_{i}^{J}+(1-r_{i}^{J})r_{i}^{A}$.

In this case,
$\langle\phi_{1}^{(3)}|\phi_{2}^{(3)}\rangle=\langle\phi_{1}^{(1)}|\phi_{2}^{(1)}\rangle\langle\phi_{1}^{(2)}|\phi_{2}^{(2)}\rangle.$
The process of calculating $r_{max}^{III}$ is similar to the proof
of $Theorem\ 5$. As a result, we can consider $r_{max}^{III}$ as
the maximum success probability with supplementary information
contained in the quantum states of one auxiliary system
$\{|\phi_{1}^{(3)}\rangle,\ |\phi_{2}^{(3)}\rangle\}$.

Furthermore, according to {\it Theorem 5}, when $P_{1}=P_{2}$, the
equality $r_{max}^{I}=r_{max}^{III}$ always holds. In other words,
there always exists a machine with one auxiliary system whose
maximum success probability of cloning is equal to that with
two auxiliary systems. \qed\\

It seems as if we could permanently increase the probability of
success. Starting from one scheme with one auxiliary system
$\{|\phi_{1}^{(1)}\rangle,\ |\phi_{2}^{(1)}\rangle\}$, we can add
second auxiliary system $\{|\phi_{1}^{(2)}\rangle,\
|\phi_{2}^{(2)}\rangle\}$ and thus increase the probability of
success. And we can always find another probabilistic cloning with
one auxiliary system $\{|\phi_{1}^{(3)}\rangle,\
|\phi_{2}^{(3)}\rangle\}$ having the same probability of success
as the previous one with two auxiliary systems. As we have now a
probabilistic cloning with one auxiliary system, we can add second
auxiliary system increasing the probability of success ones more.
Continuing this ``cyclic" argument, we can permanently increase
the probability of success.

Nevertheless, in fact, it is not like this. According to {\it
Corollary 6}, the probability increase should saturate at $1$.
Firstly, starting from one scheme with one auxiliary system
$\{|\phi_{1}^{(1)}\rangle,\ |\phi_{2}^{(1)}\rangle\}$, we can add
second auxiliary system $\{|\phi_{1}^{(2)}\rangle,\
|\phi_{2}^{(2)}\rangle\}$ satisfying the condition
$|\beta|>|\alpha|^{m-1},\ |\gamma|>|\alpha|^{m-1},\ |\beta|\neq
1,\ |\gamma|\neq 1$, and thus increase the probability of success.
Secondly, we can always find another probabilistic cloning with
one auxiliary system described by states
$\{|\phi_{1}^{(3)}\rangle,\ |\phi_{2}^{(3)}\rangle\}$, satisfying
the condition that
$\langle\phi_{1}^{(3)}|\phi_{2}^{(3)}\rangle=\langle\phi_{1}^{(1)}|\phi_{2}^{(1)}\rangle\langle\phi_{1}^{(2)}|\phi_{2}^{(2)}\rangle$,
having the same probability of success as the previous one with
two auxiliary systems. (However, we should note that the inner
product of states $|\phi_{1}^{(3)}\rangle$ and
$|\phi_{2}^{(3)}\rangle$ is less than $|\beta|$.) If we continue
this ``cyclic" argument, we can increase the probability of
success. However, notably, when the inner product of states of
auxiliary system is not more than $|\alpha|^{m-1}$, it does not
satisfy the condition of {\it Corollary 6}. So, we can not
permanently continue this ``cyclic" argument. According to {\it
Theorem 5}, when the inner product of states of auxiliary system
is not more than $|\alpha|^{m-1}$, the maximum success probability
is $1$.

\begin{center}
\textbf{4. Concluding remarks }
\end{center}

In this Letter, we considered three probabilistic cloning
protocols in terms of different communication channels between the
original party and the two supplementary parties: first, the
original party and two supplementary parties are in quantum
communication; second, the original party and the first
supplementary party are in quantum communication, but the first
supplementary party and the second one are in classical
communication; third, the original party and the first
supplementary party are in classical communication, but the two
supplementary parties are in quantum communication. In particular,
we show that, when the two states have the same {\it a priori}
probability chosen,   the best efficiency of producing $m$ copies
is achieved by the first and the third protocols. Furthermore, we
also show that, in certain circumstances, by adding an auxiliary
system, we may increase the maximum success probability of the
probabilistic cloning with supplementary information contained in
the quantum states of only {\it one} auxiliary system. However, we
always can find another probabilistic cloning with {\it one}
auxiliary system having the same success probability as that with
{\it two} auxiliary systems.

Probabilistic cloning may get precise copies with a certain
probability, so, improving the success ratio is of importance. We
hope that our results would provide some useful ideas in
preserving important quantum information, parallel storage of
quantum information in a quantum computer, and quantum
cryptography.

An interesting problem is what is the case for the two states to
be cloned having different prior probabilities chosen. As well,
novel cloning machine by Pati [15] with supplementary information
contained in the quantum states of two auxiliary systems still
merits consideration. Moreover, if the supplementary information
is given as a mixed state or we have more than two auxiliary
systems, then the probabilistic cloning devices are still worth
considering. We would like to
explore these questions in future.\\

\section*{Acknowledgement}
The authors would like to thank the anonymous referee and
Professor P.R. Holland, Editor for their invaluable comments and
suggestions that greatly help to improve the quality of this
paper.


\begin{thebibliography}{99}

\bibitem{} W.K. Wootters and W.H. Zurek, Nature 299 (1982) 802.

\bibitem{} D. Dieks, Phys. Lett. A 92 (1982) 271.

\bibitem{} H.P. Yuen, Phys. letter. A 113 (1986) 405.

\bibitem{} A.K. Pati and S.L. Braunstein, Nature 404 (2000) 164.

\bibitem{} G.M. D'riano and H.P. Yuen, Phys. Rev. Lett. 76 (1996) 2832.

\bibitem{} D. W. Qiu, Phys. Rev. A 65 (2002) 052303.

\bibitem{} H. Barnum, C.M. Caves, C.A. Fuchs, R. Jozsa and B. Schumacher, Phys. Rev. Lett. 76 (1996) 2818.

\bibitem{} M. Koashi and N. Imoto, Phys. Rev. Lett. 81 (1998) 4264.

\bibitem{} N. Gisin, G. Ribordy, W. Tittel and H. Zbinden, Rev. Mod. Phys. 74 (2002) 145.

\bibitem{} V. Scarani, S. Iblisdir, N. Gisin,  Rev. Mod. Phys.  77 (2005)
1225.

\bibitem{} V. Bu$\breve{z}$ek and M. Hillery, Phys. Rev. A 54 (1996) 1844.

\bibitem{} L.M. Duan and G.C. Guo, Phys. Lett. A 243 (1998) 261.

\bibitem{} L.M. Duan and G.C. Guo, Phys. Rev. Lett. 80 (1998) 4999.

\bibitem{} A. Chefles and S. M. Barnett, J. Phys. A 31 (1998) 10097.

\bibitem{} A. K. Pati, Phys. Rev. Lett. 83 (1999) 2849.

\bibitem{} D. Bru$\ss$, D. P. DiVincenzo, A. Ekert, C. A. Fuchs, C. Macchiavello and J. A. Smolin, Rhys. Rev. A 57 (1998) 2368.

\bibitem{} A. Chefles and S.M. Barnett, Phys. Rev. A 60 (1999) 136.

\bibitem{} R. Jozsa, IBM J. RES. DEV. 48 (2004) 79.

\bibitem{} M. Horodecki, R. Horodecki, A. Sen and U. Sen, quant-ph/0407038.

\bibitem{} K. Azuma, M. Koashi, H.Katsura, and N. Imoto, quant-ph/0604009.

\bibitem{} K. Azuma, J. Shimamura, M. Koashi and N. Imoto, Phys. Rev. A 72 (2005) 032335.

\bibitem{} D.W. Qiu, J. Phys. A 39 (2006) 5135.
Also quant-ph/0511058.

\end{thebibliography}
\end{document}